\documentstyle[psfig,floats,aps] {revtex}
\begin{document}
\newcommand{\vel}{{\bf v}}

\twocolumn[\hsize\textwidth\columnwidth\hsize\csname@twocolumnfalse\endcsname
\author{Peter B.\ Weichman$^1$ and Dean M.\ Petrich$^2$}
\address{$^1$Blackhawk Geometrics, 301 Commercial Road, Suite B, 
Golden, CO 80401 \\ $^2$Condensed Matter Physics 114-36, 
California Institute of Technology, Pasadena, CA 91125}

\title{Equilibrium solutions of the shallow water equations}
\date {{\today}}
\maketitle

\begin{abstract}
A statistical method for calculating equilibrium solutions of the 
shallow water equations, a model of essentially 2-d fluid flow
with a free surface, is described.  The model contains a competing
acoustic turbulent {\it direct} energy cascade, and a 2-d turbulent 
{\it inverse} energy cascade.  It is shown, nonetheless that, just 
as in the corresponding theory of the inviscid Euler equation, the 
infinite number of conserved quantities constrain the flow sufficiently 
to produce nontrivial large-scale vortex structures which are solutions
to a set of explicitly derived coupled nonlinear partial differential 
equations.
\end{abstract}
\vskip2pc]

\narrowtext

The evolution of a fluid from a strongly random initial condition
is generally characterized by one or more turbulent cascades of
energy to larger and/or smaller scales.  Whether energy flows
to smaller scales via a direct cascade, or to larger scales via 
an inverse cascade, is determined by a combination of conservation 
laws and phase space considerations.  Generally, if only energy is 
conserved\cite{visc} (as for 3-d Navier-Stokes turbulence\cite{MY71}), 
its flow in phase space will be globally unconstrained and will spread 
out to arbitrarily high wavenumbers, eventually draining all energy out 
of any large-scale macroscopic flows initially present.  Perhaps the
most familiar example of this is the thermodynamic equilibration of
a container of gas to a macroscopically featureless final state in 
which all energy eventually ends up as heat, i.e., microscopic 
molecular motion.  If, however, one or more additional conservation 
laws are present (as for 2-d Navier-Stokes turbulence\cite{MY71}, 
or for deep water surface gravity wave turbulence\cite{sgwaves}) 
their multiple enforcement will generally not permit \emph{both} 
conserved quantities to escape to small scales, and macroscopic 
structure, whose profile will be initial condition and boundary 
condition dependent, will survive.  An example is the equilibration 
to a \emph{rigidly rotating} final state of a gas in a \emph{cylindrical} 
container with frictionless walls.  The additional conservation of 
angular momentum along the axis of the cylinder precludes a featureless 
final state.

A long-standing problem has been the characterization of final 
states of systems with an \emph{infinite} number of conservation 
laws\cite{marsden}.  Different values of the conserved quantities 
should then produce an infinite dimensional space of final 
states\cite{foot1}.  The example of the equilibrating gas motivates 
one to postulate that the macroscopic final state, be it featureless 
or not, should be thermodynamic in character, i.e., it should be an 
\emph{equilibrium} state computable from the appropriate Hamiltonian 
using the formalism of statistical mechanics.  In \cite{MWC} this 
approach was used to produce a full characterization of the equilibrium 
states of the 2-d incompressible Euler equation (the inviscid limit of 
the 2-d Navier-Stokes equation)\cite{ergodic}, where the conserved 
quantities are the standard integrals of all powers of the vorticity.  
The equilibria were found to be characterized by a macroscopic steady 
state vorticity distribution $\omega_0({\bf r})$ obeying an explicit 
``mean field'' partial differential equation whose input parameters 
were determined by the values of the conserved quantities.

The 2-d Euler equation is the simplest of these systems in the 
sense that the incompressibility constraint $\nabla \cdot {\bf v} 
= 0$ reduces the dynamics to that of the single scalar vorticity 
field $\omega = \nabla \times \vel \equiv \partial_x v_y - \partial_y 
v_x$, and the conservation laws then provide an infinite sequence of 
global constraints on its evolution.  In this work we study a more 
complicated system of equations, the shallow water equations, an 
extension of the 2-d Euler equation that includes a free surface with 
height field $h({\bf r})$ coupled to gravity $g$.  The horizontal velocity 
${\bf v}$ now becomes compressible (with 3-d incompressibility enforced 
via $v_z = -z \nabla \cdot \vel$) but is assumed to be independent of 
the vertical coordinate $z$.  The effective 2-d dynamical equations are: 
\begin{eqnarray}
{D \vel \over D t} \equiv \partial_t \vel 
+ (\vel \cdot \nabla) \vel &=& - g \nabla h
\label{1} \\
\partial_t h + \nabla \cdot (h \vel) &=& 0,
\label{2}
\end{eqnarray}
The first equation expresses the fact that the fluid accelerates in 
response to gradients in the surface height, and the second enforces 
mass conservation, i.e., the full 3-d incompressibility.  The Euler 
equation is recovered formally when $g \to \infty$ since height 
fluctuations are then suppressed.  It is straightforward to verify 
that the ratio $\Omega \equiv \omega/h$ is convectively conserved, 
$D\Omega/Dt \equiv 0$, implying conservation of all integrals of the 
form
\begin{equation}
C_f = \int d^2r h({\bf r}) f[\Omega({\bf r})]
\label{3}
\end{equation} 
for any function $f(s)$.  These may be fully characterized by the 
function $g(\sigma)$, $-\infty < \sigma < \infty$, obtained from 
(\ref{3}) with $f(s) = \delta(\sigma - s)$, and $g(\sigma)d\sigma$ therefore 
represents the 3-d volume on which $\sigma \leq \Omega \leq \sigma+d\sigma$.  
For general $f$ one then recovers $C_f = \int d\sigma f(\sigma)g(\sigma)$.  
Note that if $\omega \equiv 0$ initially, then it must remain zero for all 
time.  Initial conditions of this type then generate (nonlinear, in general) 
wave motions \emph{only}\cite{nlwaves}.

The extension of the equilibrium theory to the shallow water equations is 
a significant advance because in addition to the usual vortical motions 
they contain acoustic wave motions\cite{foot2}.  The latter are known
\cite{acoustic} to have a direct cascade of wave energy to small scales.  
One then has the very interesting situation in which there are two 
\emph{competing} energy cascades, and the question arises as to which 
one ``wins.''  In particular, is it possible that the macroscopic vortex 
structures can ``radiate'' wave energy and disappear entirely?  We will 
show that under reasonable physical assumptions a finite fraction of the 
energy remains in large scale vortex structures, and we will derive exact 
mean field equations for the equilibrium structure.

The statistical formalism proceeds in a sequence of well defined steps.
First, the Hamiltonian corresponding to (\ref{1}) and (\ref{2}) is
\begin{equation}
{\cal H} = {1 \over 2} \int d^2 r (h {\bf v}^2 + g h^2),
\label{4}
\end{equation}
though the Poisson bracket yielding (\ref{1}) and (\ref{2}) from 
(\ref{4}) is noncanonical\cite{marsden}.  Second, the partition function 
is defined as an integral over the phase space of fields $h,{\bf v}$ 
with an appropriate statistical measure.  This so-called invariant 
measure is most easily computed if the dynamics can be expressed in terms 
of a set of variables, canonical variables being an example, for which a 
Liouville theorem is satisfied.  In this case invariant measures are
any function of the conserved integrals, with different choices corresponding
to different ensembles.  In the Euler case\cite{MWC} the field 
$\omega$ itself satisfies a Liouville theorem.  In the shallow water case 
no obvious combination of $h$, ${\bf v}$ or their derivatives meet this 
requirement.

To circumvent this problem we transform to a \emph{Lagrangian} description,
in terms of interacting infinitessimal parcels of fluid of equal 3-d volume,
for which canonical variables are easy to construct.  Thus, let ${\bf a}$
be a 2-d labeling of the system, and let ${\bf r}({\bf a},t)$ be the position 
of the parcel of fluid such that, e.g., ${\bf r}({\bf a},0) = {\bf a}$.  Since
all parcels have equal mass, the conjugate momentum is ${\bf p}({\bf a},t) 
= \dot{\bf r}({\bf a},t) = {\bf v}({\bf r}({\bf a},t),t)$.  The height field
is simply the Jacobian of the transformation between ${\bf r}$ and ${\bf a}$:
\begin{equation}
{h_0/h({\bf r}({\bf a}))} = \det ({\partial {\bf r}/\partial {\bf a}}) 
= \partial_{a_1} r_2 - \partial_{a_2} r_1,
\label{5}
\end{equation}
where $h_0$ is the overall mean height.  The Hamiltonian (\ref{4}) now takes
the form
\begin{equation}
{\cal H} = {h_0 \over 2} \int d^2a [{\bf p}({\bf a})^2 + g h({\bf a})],
\label{6}
\end{equation}
while,
\begin{eqnarray}
\omega({\bf a}) &\equiv& \nabla \times {\bf v} 
= h (\partial_{a_2} r_1 \partial_{a_1} p_1 
- \partial_{a_1} r_1 \partial_{a_2} p_1 
\nonumber \\ &&+~\partial_{a_2} r_2 \partial_{a_1} p_2 
- \partial_{a_1} r_2 \partial_{a_2} p_2) \nonumber \\
q({\bf a}) &\equiv& \nabla \cdot {\bf v} 
= h (\partial_{a_2} r_2 \partial_{a_1} p_1 
- \partial_{a_1} r_2 \partial_{a_2} p_1 
\nonumber \\ &&+~\partial_{a_2} r_1 \partial_{a_1} p_2 
- \partial_{a_1} r_1 \partial_{a_2} p_2).
\label{7}
\end{eqnarray}
It is easily verified that the Lagrangian forms of (\ref{1}) and (\ref{2})
follow from the Hamiltonian equations of motion $\dot {\bf r}({\bf a}) = 
\delta {\cal H}/\delta {\bf p}({\bf a})$ and $\dot {\bf p}({\bf a}) = 
-\delta {\cal H}/\delta {\bf r}({\bf a})$.  The Liouville theorem, which
is a statement of incompressibility of flows in phase space,
\begin{eqnarray}
&&\sum_\alpha \int d^2a [\delta \dot r_\alpha({\bf a})/\delta r_\alpha({\bf a})
+ \delta \dot p_\alpha({\bf a})/\delta p_\alpha({\bf a})] \label{8} \\
&&=\ \int d^2a [\delta^2 {\cal H}/\delta r_\alpha({\bf a}) \delta p_\alpha({\bf a})
- \delta^2 {\cal H}/\delta p_\alpha({\bf a}) \delta r_\alpha({\bf a})] = 0,
\nonumber
\end{eqnarray}
then follows immediately and implies that the correct statistical measure is 
$\rho({\cal H},\{g(\sigma)\}) \prod_{\bf a} d^2r({\bf a}) d^2p({\bf a})$.  
In the \emph{grand canonical ensemble}, which we shall adopt, the function 
$\rho$ is given by $\rho = e^{-\beta {\cal K}}$, where $\beta = 1/T$ is a
hydrodynamic ``temperature'' and
\begin{eqnarray}
{\cal K} &=& {\cal H} - \int d\sigma \mu(\sigma) g(\sigma) 
\nonumber \\
&=& {\cal H} - \int d^2r h({\bf r}) \mu[\omega({\bf r})/h({\bf r})] 
\nonumber \\ 
&=& {\cal H} - h_0 \int d^2a \mu[\omega({\bf a})/h({\bf a})] 
\label{9}
\end{eqnarray}
in which $\mu(\sigma)$ is a chemical potential that couples to each level
$\omega({\bf r})/h({\bf r}) = \sigma$.  The partition function is now
defined by
\begin{equation}
Z[\beta,\{\mu(\sigma)\}] = {1 \over N!} \prod_{\bf a}
\int d^2r({\bf a}) \int d^2p({\bf a}) e^{-\beta {\cal K}},
\label{10}
\end{equation}
where $N \to \infty$ is the number of fluid parcels and $N!$ is the usual 
classical delabeling factor.  The thermodynamic averages of the conserved 
quantities  are now obtained in the usual fashion as derivatives with respect
to the chemical potentials, $\langle g(\sigma) \rangle = T \delta \ln(Z)/
\delta \mu(\sigma)$. 

One would now like to transform the integration in (\ref{10}) back to
physical Eulerian variables.  The key observation is that, from (\ref{7}),
$\Omega \equiv \omega/h$ and $Q \equiv q/h$ are \emph{linear} in ${\bf p}$.  
Therefore, one may formally invert this relationship to obtain $\prod_{\bf a} 
d^2p({\bf a}) = \prod_{\bf a} dQ({\bf a}) d\Omega({\bf a})
J[h]$, where, due to the particle relabeling symmetry (both $\nabla \cdot 
{\bf v}$ and $\nabla \times {\bf v}$ depend only on ${\bf r}$ and are then 
clearly invariant under any permutation of the labels ${\bf a}$), the 
Jacobian $J$ is a functional of the height field $h({\bf a})$ \emph{alone}.  
The exact form of $J$ will turn out to be unimportant.  Similarly, 
$(1/N!) \prod_{\bf a} \int d^2r({\bf a}) = \prod_{\bf a} \int 
dh({\bf a}) I[h]$, where $I[h]$ is another Jacobian.  The $1/N!$ 
factor precisely removes the relabeling symmetry that, in particular, 
leaves the height field invariant.  Finally, we replace the label ${\bf a}$
by the actual position ${\bf r}$, in which the equal volume restriction
on each fluid parcel implies that the infinitesimal area of each parcel 
must be determined by $dV = h({\bf r}) d^2r = constant$.  Thus:
\begin{eqnarray}
&&{1 \over N!} \prod_{\bf a} \int d^2r({\bf a}) \int d^2p({\bf a})
\nonumber \\
&&~~~~~~~~~~=~\prod_{\bf r} \int dh({\bf r}) {\cal J}[h] 
\int d\Omega({\bf r}) \int dQ({\bf r}),
\label{11}
\end{eqnarray}
in which ${\cal J}[h] = I[h]J[h]$, and the mesh over which the label 
${\bf r}$ runs is \emph{nonuniform} and \emph{changes} with each 
realization of the height field $h$.

The statistical operator ${\cal K}$ must also be expressed in terms of
$Q,\Omega,h$.  Only for the kinetic energy $T = \int d^2r h {\bf v}^2$ 
does this require some nontrivial manipulations.  Let the current ${\bf j} 
\equiv h {\bf v}$ be decomposed in the form ${\bf j} = \nabla \times \psi 
- \nabla \phi$.  One obtains then
\begin{equation}
\left(\begin{array}{c} h \Omega \\ h Q \end{array} \right)
= \left(\begin{array}{cc} \nabla \times {1 \over h} \nabla \times &
- \nabla \times {1 \over h} \nabla \\
\nabla \cdot {1 \over h} \nabla \times & - \nabla \cdot {1 \over h} \nabla
\end{array} \right)
\left(\begin{array}{c} \psi \\ \phi \end{array} \right).
\label{12}
\end{equation}
The $2\times 2$ matrix operator, which we shall denote ${\cal L}_h$, appearing
on the right hand side of (\ref{12}) is self adjoint and positive definite,
and therefore possesses an inverse, i.e., a $2 \times 2$ matrix Green function
${\cal G}_h({\bf r},{\bf r}')$ satisfying ${\cal L}_h {\cal G}_h({\bf r},{\bf r}') 
= \openone \delta({\bf r}-{\bf r}')$.  An explicit form for ${\cal G}_h$ will not 
be needed.  The kinetic energy is then $T = {1 \over 2} \int d^2r {\bf j} \cdot 
{\bf v} = \int d^2r h(\psi \Omega + \phi Q)$, i.e.,
\begin{equation}
T = \int d^2r h({\bf r}) \int d^2r' h({\bf r}') 
\left(\begin{array}{c} \Omega({\bf r}) \\ Q({\bf r}) \end{array} \right) 
{\cal G}_h({\bf r},{\bf r}')
\left(\begin{array}{c} \Omega({\bf r}') \\ Q({\bf r}') \end{array} \right)
\label{13}
\end{equation}
and the complete statistical operator is
\begin{equation}
{\cal K} = T + \int d^2r h({\bf r}) \left\{{1 \over 2}g h({\bf r}) 
- \mu[\Omega({\bf r})] \right\}
\label{14}
\end{equation}
The appearance of the factors $h({\bf r})$ and $h({\bf r}')$ is crucial here
because, as discussed above, $dV = h({\bf r}) d^2r$ and $dV' = h({\bf r}')d^2r'$
are both uniform for each given statistical mesh.

We finally come to the evaluation of the partition function itself.  This
is accomplished with the use of the \emph{Kac-Hubbard-Stratanovich} (KHS)
transformation, which in discrete form reads for any positive definite matrix
${\bf A}$,
\begin{equation}
e^{{1\over 2} \sum_{i,j} y_i A_{ij} y_j} =
{1 \over {\cal N}} \prod_i \int_{-\infty}^{\infty} d\zeta_i
e^{-{1 \over 2} \sum_{i,j} \zeta_i A^{-1}_{ij} \zeta_j
-\sum_i \zeta_i \cdot y_i},
\label{15}
\end{equation}
where $y_i$ and $\zeta_i$ may be vectors, and the normalization is ${\cal N} 
= \sqrt{\det(2\pi {\bf A})}$.  This identity follows by completing the square 
on the right hand side and performing the remaining Gaussian integral.  We 
apply it to the discretized version of (\ref{10}) and (\ref{11}) with finite 
$dV$, and the identifications $A_{ij} = -\beta^{-1}{\cal G}({\bf r}_i,{\bf r}_i)$
\cite{foot3}, $y_i = \beta dV [\Omega({\bf x}_i), \Phi_i({\bf x}_i)]$ and we 
introduce the notation $\zeta_i = [\Psi({\bf x}_i),\Phi({\bf x}_i)]$.  The 
continuum limit $dV \to 0$ will be taken at the end.  The partition function 
is now
\begin{equation}
Z = \prod_i \int dh_i {{\cal J}[h] \over {\cal N}[h]} 
\int d\Psi_i d\Phi_i \int dQ_i d\Omega_i
e^{\beta {\tilde {\cal F}}},
\label{16}
\end{equation}
where
\begin{eqnarray}
{\tilde {\cal F}} &=& dV \sum_{i,j} \left(\begin{array}{c} \Psi_i \\ \Phi_i 
\end{array} \right) [{\cal L}_h]_{ij} 
\left(\begin{array}{c} \Psi_j \\ \Phi_j \end{array} \right) \nonumber \\
&&-~dV \sum_i [\Omega_i \Psi_i + Q_i \Phi_i - \mu(\Omega_i)],
\label{17}
\end{eqnarray}
in which $[{\cal L}_h]_{ij}$ is an appropriate discretization of the
differential operator ${\cal L}_h$.  Notice that the inverse of ${\cal G}_h$ 
has led to the reappearance of the local differential operator ${\cal L}_h$.  

At the expense, then of introducing the new fields $\Psi$, $\Phi$ we have 
succeeded in producing a purely \emph{local} action in which the integration 
over $\Omega_i$, $Q_i$ can be performed independently for each $i$ (for given 
fixed field $h$).  However, we now arrive at a problem whose physical origin,
as we shall see, lies precisely in the direct cascade of wave energy.  Thus,
the chemical potential function $\mu(\sigma)$ controls convergence of the 
Laplace transform-type integral
\begin{equation}
e^{\bar \beta W[\Psi_i]} \equiv \int_{-\infty}^\infty d\Omega_i
e^{-\bar \beta [\Omega_i \Psi_i - \mu(\Omega_i)]},
\label{18}
\end{equation} 
where $\bar \beta \equiv \beta dV$ corresponds to a rescaled hydrodynamic
temperature $\bar T = T dV$ which is assumed to remain \emph{finite} as
$dV \to 0$---the object of this choice is to obtain the correct control
parameter for nontrivial hydrodynamic equilibria in the continuum limit
that, as we shall see, yields a nontrivial balance between energy and
entropy contributions to the final free energy\cite{MWC}.  However, 
there is no corresponding chemical potential controlling $Q_i$ and the 
corresponding integral does not converge.  Recalling that $Q = (1/h) 
\nabla \cdot {\bf v}$, unboundedness of $Q$ reflects unboundedness of 
small-scale gradients in the compressional part of ${\bf v}$ and in 
$h$\cite{foot4}.  Thus, taken literally, the direct cascade of wave 
energy leads to arbitrarily small scale fluctuations of the fluid 
surface that remain of fixed amplitude, i.e., a kind of foam of fixed 
thickness.  Physically, of course, such small scale motions are rapidly 
dissipated by processes that violate the approximations used to derive 
the shallow water equations, e.g., by some combination of viscosity and 
wave breaking\cite{foot5}.  This leads to the following physically 
motivated assumption:  dissipative processes that suppress wave motions 
lead to the interpretation $\int dQ_i \exp(\bar \beta Q_i \Phi_i)
\to \delta(\bar\beta \Phi_i)$, i.e., to the vanishing of $\Phi_i$.

With $\Phi_i \equiv 0$, only the $(1,1)$ component of ${\cal L}_h$
contributes, and in the continuum limit $dV \to 0$ the partition
function becomes
\begin{equation}
Z = \prod_{\bf r} \int dh({\bf r}) {{\cal J}[h] \over {\cal L}[h]}
\int d\Psi({\bf r}) e^{-\beta {\cal F}[h,\Psi]},
\label{19}
\end{equation}
where the Free energy functional is
\begin{equation}
{\cal F} = - \int d^2r \left[{(\nabla \Psi)^2 \over 2h}
- {1 \over 2} g h^2 + h W[\Psi] \right].   
\label{20}
\end{equation}
The key observation now is that $\beta = \bar \beta/dV \to \infty$
in the continuum limit.  Thus, mean field theory becomes \emph{exact}
and equilibrium solutions are given by \emph{extrema} of ${\cal F}$.
This is why the integration over the field $h({\bf r})$, with its
unknown Jacobian, is ultimately irrelevant.  The underlying assumption
is only that the Jacobian is smooth, or at least less singular than
$e^{-\beta {\cal F}}$, in the neighborhood of the extremum in the 
continuum limit.

The extremum conditions $\delta {\cal F}/\delta \Psi({\bf x}) =
0 = \delta {\cal F}/\delta h({\bf x})$ yield then the mean field 
equations
\begin{eqnarray}
\nabla \cdot \left[{1 \over h({\bf r})} \nabla \Psi \right]
&=& h({\bf r}) W'[\Psi({\bf r})] 
\label{21} \\
{[\nabla \Psi({\bf r})]^2 \over 2h({\bf r})^2} &=&
W[\Psi ({\bf r})] - g h({\bf r}),
\label{22}
\end{eqnarray}
By adding a source term $\int d^2r h({\bf r}) \tau({\bf r}) 
\Omega({\bf r})$ to ${\cal K}$, which serves only to replace $\Psi$
by $\Psi - \tau$ inside $W$, one may compute the equilibrium
average $\langle \Omega({\bf r}) \rangle = [\delta {\cal F}/\delta 
\tau({\bf r})]_{\tau \equiv 0} = -h^{-1} \nabla \cdot (h^{-1} \nabla 
\Psi)$.  It follows then that $\langle {\bf j} \rangle = \nabla \times 
\Psi$, so that $\Psi$ is the \emph{stream function} associated with 
the equilibrium current.  Equation (\ref{21}) is in fact equivalent 
to $\Omega = -W'(\Psi)$, which guarantees that this is a true 
equilibrium solution satisfying ${\dot \Omega} = 0$, and equation
(\ref{22}) is equivalent to Bernoulli's theorem since it can be 
rewritten as $(1/2) {\bf v}^2 + g h = W(\Psi)$.

As a simple example, in the case where $\Omega = \sigma_0$ over 
half the 3-d volume of the fluid and $\Omega = 0$ on the other 
half, the chemical potential takes the form $e^{\bar \beta \mu(\sigma)} 
= e^{\bar \beta \mu_0} \delta(\sigma) + e^{\bar \beta \mu_1} 
\delta(\sigma-\sigma_0)$, and therefore by (\ref{18}) 
$e^{{\bar \beta} W(s)} =  e^{{\bar \beta}\mu_0} + 
e^{{\bar \beta}(\mu_1 - \sigma_0 s)}$.  Extensive numerical 
solutions for the Euler equilibria exist for this ``two-level'' 
system as a function of $\beta$ and $\mu_1-\mu_0$.~\cite{MWC,ergodic}.  
In preliminary numerical work, we find that the shallow water equilibria 
generated by (\ref{22}) have very similar structure (with, for example,
vorticity moving from the walls toward the center of the system as
$\beta$ decreases from positive to negative values), while the height 
field basically covaries with the vorticity in order to maintain 
hydrostatic balance.  Details of this work will be presented 
elsewhere.

The techniques presented in this paper can be used to generate equilibrium
equations for a number of other systems with an infinite number of conserved
integrals\cite{marsden}.  The key insight presented here is that whenever such 
a system contains simultaneous direct and inverse energy cascades, the long
time dynamics becomes very singular and additional physically motivated 
assumptions must be made in order to derive sensible equilibria.  Our 
assumption, that dissipation acts to suppress the forward cascading degrees
of freedom with negligible effect on the macroscopic state, presumably
depends on the smoothness of the initial condition.  Comparisons with
detailed numerical simulations will be required to evaluate such effects.

{\bf Note added:}  After completion of this work we became aware of
an e-print \cite{CS} where equations equivalent to (\ref{22}) are 
derived from a phenomenological maximum entropy theory.  No statistical
mechanical derivation is given, nor is the interaction between wave and 
vortical motions and the effects of waves on equilibration discussed.


\begin{references}
\bibitem{visc} Real fluids are always viscous, but models of turbulence
generally concern themselves with the ``inertial range'' where loss of
energy due to viscous damping is small compared to that due to the
cascade process, and an energy conserving model is appropriate.
\bibitem{foot1} Study of these states is partly motivated by the
constraints they place on simulations of turbulent flow, e.g., the 
degree to which they properly preserve the conservation laws.
\bibitem{foot5} Dissipation processes also act on the vortex structures,
but the conservation laws guarantee that microscopic fluctuations in $\Omega$, 
unlike $Q$, remain finite and hence will be dissipated much less strongly.
\end{references}
\end{document}